\documentclass[lettersize,journal]{IEEEtran}

\usepackage{amsmath}
\usepackage{amsfonts}
\usepackage{amssymb}

\usepackage{algorithm}
\usepackage{algpseudocode}

\usepackage{array}
\usepackage{booktabs}
\usepackage{enumitem}

\usepackage{graphicx}
\usepackage[caption=false, font=footnotesize]{subfig}
\usepackage{stfloats}

\usepackage{textcomp}
\usepackage{color}
\usepackage[svgnames]{xcolor}
\usepackage{soul}
\sethlcolor{cyan!30}
\usepackage{url}
\usepackage{balance}
\usepackage{cite}
\usepackage[hidelinks]{hyperref}
\usepackage{titlesec}
\allowdisplaybreaks

\titlespacing*{\section}
{0pt}{1.5ex plus 1ex minus .2ex}{0.8ex}

\titlespacing*{\subsection}
{0pt}{1.2ex plus 0.8ex minus .2ex}{0.6ex}

\begin{document}

\title{Traffic-Aware Microgrid Planning for Dynamic Wireless Electric Vehicle Charging Roadways}

\author{Dipanjan Ghose, Junjie Qin, and S Sivaranjani}

\maketitle

\begin{abstract}
Dynamic wireless charging (DWC) is an emerging technology that has the potential to reduce charging downtime and on-board battery  size, particularly in heavy-duty electric vehicles (EVs). 
However, its spatiotemporal, dynamic, high-power demands pose challenges for power system operations. Since DWC demand depends on traffic characteristics such as speed, density, and dwell time, effective infrastructure planning must account for the coupling between traffic behavior and EV energy consumption. In this paper, we propose a novel traffic-aware microgrid planning framework for DWC. First, we use the macroscopic cell transmission model to estimate spatio-temporal EV charging demand along DWC corridors and  integrate this demand into an AC optimal power flow formulation to design a supporting microgrid. Our framework explicitly links traffic patterns with energy demand and demonstrates that traffic-aware microgrid planning yields significantly lower system costs than worst-case traffic-based approaches. We demonstrate the performance of our model on a segment of I-210W in California under a wide range of traffic conditions.
\end{abstract}

\begin{IEEEkeywords}
Dynamic Wireless Charging, Electric Vehicles, Traffic Flow, Microgrid Planning
\end{IEEEkeywords}

\section{Introduction}\label{sec: Introduction}

Dynamic wireless charging (DWC) on electrified roadways is an emerging technology that allows electric vehicles (EVs) to charge on-the-go, reducing charging downtime and range anxiety, while significantly decreasing on-board battery size requirements and consequently vehicle cost \cite{lukicCuttingCordStatic2013,panchalReviewStaticDynamic2018a,yeJointPlanningDynamic2024}. DWC uses existing road infrastructure and allows multiple EVs to charge simultaneously, thus alleviating land acquisition concerns and promoting efficient use of roadway infrastructure. Recently, DWC has attracted increasing attention worldwide, with multiple pilot projects currently underway \cite{indotWirelessElectricVehicle2021, lozanovaDynamicWirelessCharging2025, galigekere2021high, electreon2023trondheim}. 

Despite these early successes, large-scale DWC deployment faces several unresolved challenges, foremost among them the planning of a supporting power infrastructure. Unlike static EV charging demands~\cite{fonsecaEVLearnExtendingCityLearn2025,ghoseStudyingPresentFuture2023,el2022impact}, DWC requires high-power, dynamic, uninterrupted energy transfer to vehicles in motion under continuously varying supply and demand conditions. The charging demand is governed by real-time traffic flow, leading to spatiotemporal demand variability driven by vehicle density, instantaneous power transfer requirements, charging lane usage, and driving characteristics such as braking and inter-vehicular spacing. The resulting power demand profile is highly dynamic, non-uniform, and difficult to predict, making DWC a paradigm in which transportation networks and power systems are closely coupled. 

This close coupling raises several questions critical to power infrastructure planning: {how do we model dynamic traffic-driven energy demand from DWC? How do these  spatio-temporally varying demand patterns influence the design and operation of the supporting power grid? How can we reliably serve DWC demand under varying traffic conditions?} Prior studies~\cite{wangOptimalControlStrategy2024, Liu2024, majidiDynamicInmotionWireless2024,dong2024real} address these questions using microscopic (vehicle-level) models that track individual EV routes, driving patterns, and charging behavior to estimate charging demand and design charging control mechanisms. However, microscopic traffic modeling is computationally intensive, intractable for large-scale DWC deployments with limited data availability, and raises privacy concerns. Recent works on grid planning for DWC also aim to coordinate charging demand with grid operations \cite{Karakitsios2016, sunOptimizingDynamicWireless2024, majidiSpatioTemporalImpactAnalysis2024, aduamaOptimizingQuasidynamicWireless2024}; however, these works either rely on simplified grid models that do not fully capture grid operational constraints, employ control strategies that  influence traffic behavior, or are restricted to vehicle classes that satisfy predictable routing patterns, such as buses and taxis. In this context, there is a need for systematic approaches for DWC grid planning integrating realistic but tractable models of traffic dynamics and grid constraints. 

In this paper, we bridge this gap by proposing a \textit{traffic-aware grid planning framework} that couples two components: (i) a macroscopic traffic model to estimate spatiotemporal DWC energy demands  derived from the cell transmission model (CTM) of traffic flows \cite{daganzo1}, and (ii) a planning problem that takes into account these macroscopic traffic demand patterns to optimally size and site a microgrid with solar power and energy storage (ES), while accounting for operational constraints like voltage limits, power flows, and capacity constraints. On the traffic modeling side, we use a macroscopic representation of traffic dynamics 
to efficiently estimate DWC demands without vehicle-level monitoring or case-specific models \cite{ferraraFirstOrderMacroscopicTraffic2018, zaheerDynamicEVCharging2017, aungDynamicTrafficCongestion2021, wangReliableDynamicWireless2024, elmeligyOptimalPlanningDynamic2024,
ngoOptimalPositioningDynamic2020}.
On the grid side, we focus on microgrids, as they are the preferred architecture in DWC pilots such as the planned Pennsylvania Turnpike freight charging corridors \cite{Jack2024} due to their operational flexibility and relative independence from the external grid  \cite{ALRUBAIE20255590}. By embedding a traffic-dependent characterization of DWC loads in the supporting microgrid design problem, our approach minimizes the cost of DWC grid infrastructure while ensuring reliable operation across a variety of real-world traffic conditions. 
Our main contributions are summarized as follows:
\begin{enumerate}[label=\alph*)]
\item We develop a novel traffic-aware dynamic energy demand modeling framework that leverages macroscopic traffic flow dynamics to estimate spatiotemporal EV charging demand across DWC corridors without relying on vehicle-level characteristics or driving behaviors.

\item We couple this traffic energy demand model with a planning problem built on a second-order cone programming (SOCP) relaxation of the AC-OPF to optimize the sizing and siting of a microgrid for DWC while ensuring voltage support and satisfaction of grid constraints.

\item We construct realistic traffic  reflecting a range of conditions from normal traffic variability to traffic incidents to rigorously assess the reliability of the microgrid under diverse demand conditions.

\end{enumerate}

We demonstrate our framework through simulation on a 14-mile segment of the I-210 W highway in Los Angeles under a variety of traffic conditions \cite{CTMSIMInteractiveFreeway2025}.  Our results show that traffic-aware DWC demand modeling enables more efficient infrastructure planning by reducing grid oversizing and lowering capital and operational costs. To the best of our knowledge, this work is among the first to integrate macroscopic traffic flow models with grid planning for DWC infrastructure.


This paper is organized as follows. In Section \ref{sec: Motivating Example} , we begin by presenting motivating example that illustrates the need for traffic-aware DWC demand modeling. We then outline the problem formulation, traffic modeling, and microgrid planning framework in Section~\ref{sec: Methodology}. We describe the I210-W case study in  Section~\ref{sec: Case Study - Set-Up} and present the simulation results in Section~\ref{sec: Case Study - Results}. Finally, we discuss our key findings in Section~\ref{sec: Discussion}.

\section{Motivating Example}\label{sec: Motivating Example}
\begin{figure}[t]
    \centering
    \includegraphics[width=\linewidth]{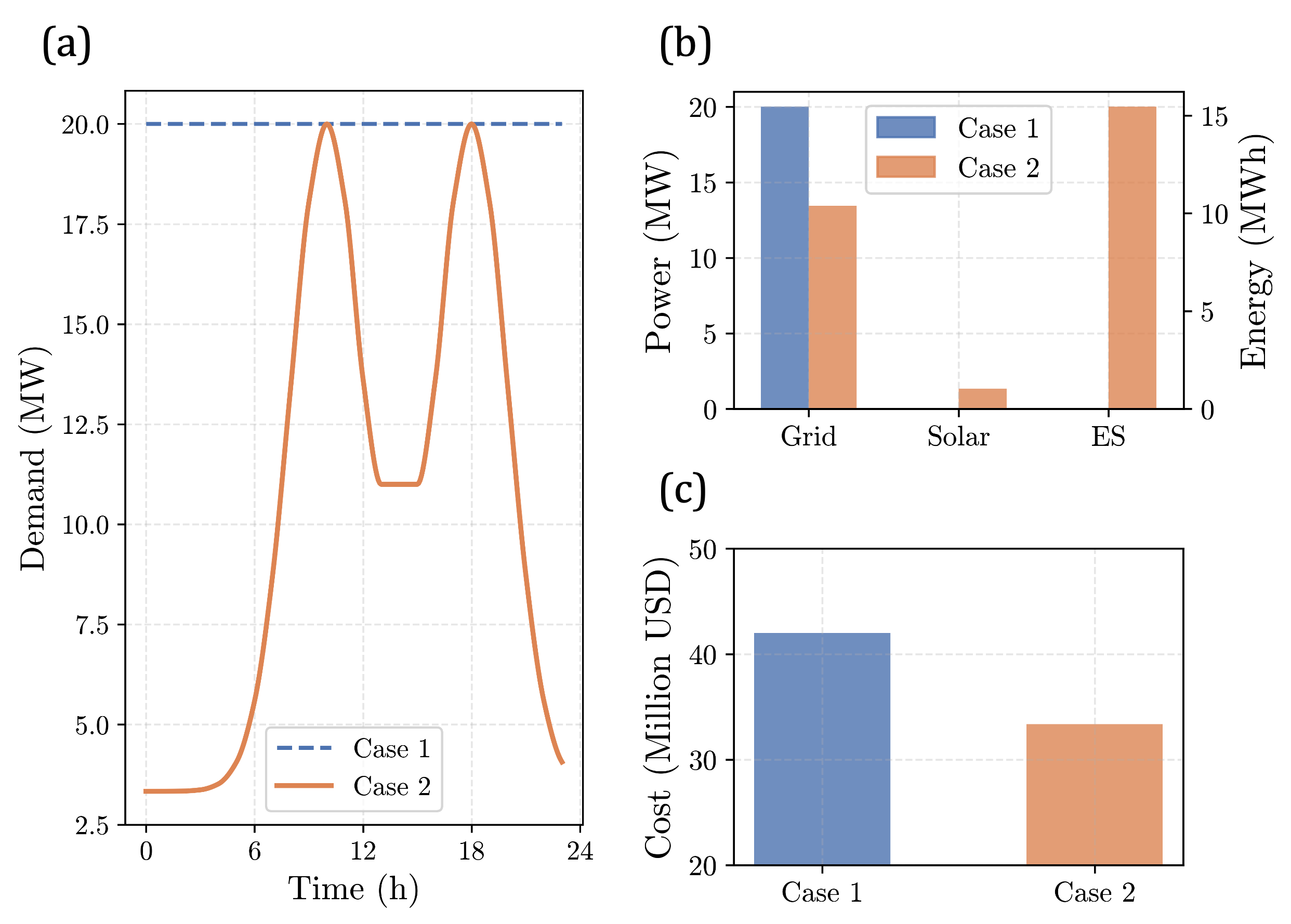}
    \caption{Results for the motivating example showing: (a) the demand curves for cases 1 and 2, with morning and evening traffic peaks, (b) the corresponding sizing of grid components (grid capacity, solar, and ES) for cases 1 and 2, and (c) a comparison of infrastructure costs, demonstrating significant savings when traffic-aware microgrid planning is considered. Note that for (b), the ES units are in MWh, while the grid and solar capacities are in MW.}
    \label{fig:motivation_combined}
    \vspace{-0.5cm}
\end{figure}

We begin with a stylized version of the DWC microgrid planning problem  to  illustrate that incorporating traffic dynamics in DWC demand modeling, rather than designing systems for continuous worst-case demand, can reduce system overdesign and associated infrastructure costs. Note that the generalized and detailed formulation of the DWC microgrid planning problem is introduced later in Section~\ref{sec: Methodology}. The microgrid consists of solar generation, energy storage (ES), and external grid coupling with capacities $P^{\text{S}}$ (MW), $P^{\text{G}}$ (MW), and $E^{\text{ES}}$ (MWh), respectively, with corresponding construction and operation costs of $C^{\text{S}}\! =\! 1{,}000{,}000$ USD/MW, $C^{\text{G}} \! =\! 2{,}100{,}000$ USD/MW, and $C^{\text{ES}}\! = \!246{,}000$ USD/MWh (see Section~\ref{subsec: Dataset Description}). The objective is to minimize the microgrid capital cost required while ensuring that it always meets the DWC demand, formulated as the following problem:
\begin{subequations}\label{eq:motivating_optimization}
\begin{align}
\min_{P^{\text{S}}, P^{\text{G}}, E^{\text{ES}}} 
& C^{\text{S}} P^{\text{S}}
+ C^{\text{G}} P^{\text{G}}
+ C^{\text{ES}} E^{\text{ES}} \label{eq:motivating_obj} \\[4pt]
\text{s.t.} \quad 
& p_t^{\text{g}} + \gamma_t P^{\text{S}} + p_t^{\text{ES}} = p_t^{\text{d}}, 
&& \forall t \label{eq:motivating_balance} \\
& 0 \le p_t^{\text{g}} \le P^{\text{G}}, 
&& \forall t \label{eq:motivating_gridcap} \\
& e_t = e_{t-1} - \Delta t\, p_t^{\text{ES}}, \quad e_0 = e_T,
&& \forall t \label{eq:motivating_storage_dyn} \\
& 0 \le e_t \le E^{\text{ES}}, 
&& \forall t \label{eq:motivating_storage_cap} \\
& P^{\text{S}} \ge 0, \; P^{\text{G}} \ge 0, \; E^{\text{ES}} \ge 0.
\label{eq:motivating_nonneg}
\end{align}
\end{subequations}
Here, \eqref{eq:motivating_balance} represents the power balance at time step $t \in \{0, 1, \dots T\}$ between demand $p_t^{\text{d}}$ and supply from the external grid $p_t^{\text{g}}$, energy storage ($p_t^{\text{ES}}$), and solar $\gamma_t P^{\text{S}}$, with $\gamma_t \in [0,1]$ denoting the fractional solar generation peaking at daytime. The constraint \eqref{eq:motivating_storage_dyn} defines the ES dynamics, with $e_t$ representing the stored energy at time $t$. Considering an ideal ES, $p_t^{\text{ES}}$ can be positive or negative, depending on whether the ES is discharging or charging. 
Both $p_t^{\text{g}}$ and $p_t^{\text{ES}}$ are limited by their respective planned capacities \eqref{eq:motivating_gridcap} and \eqref{eq:motivating_storage_cap}, and all decision variables are non-negative as in \eqref{eq:motivating_nonneg}. With this setting, we compare the microgrid parameters obtained by solving \eqref{eq:motivating_optimization} under two DWC demand cases,  as summarized below and illustrated in Fig. \ref{fig:motivation_combined}(a).
\begin{itemize}
\item \textbf{Case 1:} DWC demand is modeled as a constant, that is, $\forall t, p_t^{\text{d}} = 20$ MW, representing worst-case power demand at all times, which is a common assumption in DWC grid infrastructure planning studies.
\item \textbf{Case 2:} DWC demand follows varying traffic patterns that reflect realistic traffic behavior, peaking during morning and evening rush hours and reducing at late-night periods, with the peak demand remaining 20 MW.
\end{itemize}
We  solve  problem~\eqref{eq:motivating_optimization}  over a time horizon of 24-hours, that is, with $T=23$, with the resulting microgrid parameters and associated system costs shown in Fig.~\ref{fig:motivation_combined}(b) and Fig.~\ref{fig:motivation_combined}(c) respectively. The results demonstrate that incorporating traffic-aware demand in the planning problem enables the use of flexible grid resources, prevents system overestimation, and lowers system costs, with Case 1 costs being 27\% more than Case 2. This is due to the fact that traffic-aware DWC demand modeling allows us to leverage the temporal flexibility of ES and solar generation to reduce system costs, wherein the ES charges during reduced demand periods and discharges during demand peaks, reducing reliance on fixed generation, which is costly to construct and operate.
In contrast, the demand model in Case 1 lacks such flexibility, as the system is designed to always meet peak demand, which occurs only for brief intervals, resulting in system overdesign and increased reliance on external grid power. 
In deployments involving multiple charging corridors and longer planning horizons, such overdesigns can translate into infrastructure costs in the order of millions of dollars. These observations motivate the remainder of the paper, which examines the coupling between traffic dynamics and energy demand and its implications for DWC grid planning with flexible resources such as solar generation and ES.

\section{Traffic-Aware Grid Planning Framework}\label{sec: Methodology}
We propose a traffic-aware microgrid planning framework that (i) integrates macroscopic traffic flow modeling to obtain spatiotemporal traffic states and estimate the resulting dynamic energy demand, (ii) incorporates this demand into an AC-OPF-based microgrid planning problem. The microgrid comprises solar generation, ES, and an interconnection with the external grid. The planning problem accounts for lifetime capital and operational costs, as well as voltage limits, power balance, and line capacity constraints. Longer-term factors such as EV growth and highway expansion are not considered in this study, as the focus is on the immediate impact of traffic flow on energy demand and microgrid planning.

\subsection{Cell Transmission Model}\label{subsec: Cell Transmission Model}

The cell transmission model (CTM), introduced by Daganzo in 1994~\cite{daganzo1,daganzo2}, stands out as one of the most widely adopted frameworks for modeling macroscopic traffic flow across complex networks~\cite{alimardani2021}. We adopt the CTM for our setting as its discrete, computationally efficient formulation and its ability to capture key traffic phenomena such as congestion, shockwaves, and spillback~\cite{munoz2001} makes it well suited to capture essential dynamics of traffic flows that are expected to influence DWC grid demands, without resorting to vehicle-level modeling. In the CTM, the highway is segmented into a series of cells, and time is divided into discrete intervals. The length of each cell is defined as the distance a vehicle travels in one time step under free-flow conditions, ensuring that vehicles within a cell advance to the next cell at each time step $t$~\cite{daganzo1, munozMethodologicalCalibrationCell2004}. Individual vehicle characteristics, such as velocity, are treated as constants within each cell.

Considering a linear highway segment, we denote the vehicle density in each cell $i \in \{1,2,\ldots,\mathcal{K}\}$ at time step $t$ by $\rho_{i,t}$, representing the number of vehicles per unit distance in the cell, where $\mathcal{K}$ denotes the total number of cells in the roadway. 
Each cell $i$ is then assigned piecewise-affine sending and receiving flow functions, denoted by $S_{i,t}$ and $R_{i,t}$ respectively, defined as follows~\cite{siva2017}:
\begin{align}
S_{i,t} &= \min \{ M_i, \, v^{\text{ff}}_i \, \rho_{i,t} \}, \label{eq:Si} \\
R_{i,t} &= \min \{ M_i, \, {v}^{\text{c}}_i (\rho^{\text{jam}}_i - \rho_{i,t}) \}, \label{eq:Ri}
\end{align}
where $v^{\text{ff}}_i \! \in \! \mathbb{R}^+$ and ${v}^{\text{c}}_i \! \in \! \mathbb{R}^+$ denote the free-flow and congested speeds of cell $i$, while $\rho^{\text{jam}}_i$ and $M_i$ represent its jamming density and maximum outflow at time $t$. Now, the system state is defined as the $N$-vector $\boldsymbol{\rho}_t$, obtained by stacking the density in each cell, that is, $\boldsymbol{\rho}_t \! \triangleq \! \begin{bmatrix}\rho_t^1 \dots \rho_t^N\end{bmatrix}^T$. Then, we can write $\boldsymbol{\rho}_t \! \in \! \mathcal{D}, \, \forall t \! \in \! \mathbb{Z}_{\ge 0}$ where $\mathcal{D} \!=\! \prod_{i=1}^{N} [0, \rho^{\text{jam}}_i]$. For every cell $i$, we define the inflow of traffic as:
\begin{equation}
Q_{i}^{\text{in}}(\boldsymbol{\rho}_t) =
\min \{ S_{(i-1),t}, R_{i,t} \}.
\label{eq:qin}
\end{equation}
Similarly, the outflow of traffic from the cell $i$ will be:
\begin{equation}
Q_{i}^{\text{out}}(\boldsymbol{\rho}_t) =
\min \{ S_{i,t}, R_{(i+1),t} \}.
\label{eq:qout}
\end{equation}
On- and off-ramp flows (including turning ratios) can be incorporated by suitably modifying the inflow and outflow functions in \eqref{eq:qin}-\eqref{eq:qout}. We refer the reader to \cite{daganzo2} for details. 

Finally, we define the update equation for vehicle density in each cell $i$ at  time step $t$ to define the CTM that characterizes macroscopic traffic flow dynamics as:
\begin{equation}\label{eq:CTM Update}
\rho_{i,(t+1)} = \rho_{i,t} + Q_{i}^{\text{in}}(\boldsymbol{\rho}_t) - Q_{i}^{\text{out}}(\boldsymbol{\rho}_t), \quad i \in \{1,2,\ldots, \mathcal{K}\}.
\end{equation}
The CTM in \eqref{eq:Si}-\eqref{eq:CTM Update} enables us to compute traffic densities at each location in the highway and each time, which we will leverage to estimate DWC demands. 

\subsection{Dynamic Energy Consumption Model}\label{subsec: Dynamic Energy Consumption Model}
We now develop a dynamic energy consumption model that translates traffic dynamics, such as vehicle flows and densities obtained from the CTM, into the corresponding power demand, thereby capturing the tight coupling between traffic flows and grid loads. It is important to note that the power demand on a DWC corridor does not scale proportionally with traffic density; rather, it is nonlinearly related to traffic flows through congestion-dependent dynamics as modeled in \eqref{eq:Si}-\eqref{eq:CTM Update}. Accordingly, this nonlinear relationship between traffic flow and density must be accounted for when modeling the resulting charging demand.

The power consumed by an EV to drive a certain distance depends on several vehicular parameters, such as mass, frontal area, speed, and efficiency, as well as physical factors like air resistance and rolling resistance of the wheels. In general, the driving power  $p^{\text{drive}}_{i,t}$ for each vehicle in cell $i$ at time step $t$ is calculated using the following relation~\cite{haddad2022}:
\begin{equation}
p^{\text{drive}}_{i,t} = \frac{1}{\eta^{\text{drive}}} \cdot \left( \frac{1}{2} c^{\text{d}} A \rho^{\text{air}} {(v_{i,t})}^3 + c^{\text{rr}} m g v_{i,t} \right),
\label{eqn: dynamic energy consumption}
\end{equation}
where 
 $v_{i,t}$ is the average velocity obtained for the cell $i$ at time step $t$ from the CTM,
 $\eta^{\text{drive}}$ is the efficiency of the electric drivetrain,
 $c^{\text{d}}$ is the drag coefficient,
 $c^{\text{rr}}$ is the rolling resistance coefficient,
 $A$ is the average frontal area of the EV,
 $\rho^{\text{air}}$ is the air density,
 $m$ is the average mass of the EV, and
 $g$ is the acceleration due to gravity. 

To determine the actual power required from the DWC coils, we also account for auxiliary power consumption, denoted by $p^{\text{aux}}$ which includes factors such as air conditioning.
Then, the total energy consumption for a representative vehicle in cell $i$ with average speed $v_{i,t}$ at time step $t$ is given by:
\begin{equation}
e^{\text{drive}}_{i,t} = \left( p^{\text{drive}}_{i,t} + p^{\text{aux}} \right) \cdot \Delta{t}.
\end{equation}
To scale this up to macroscopic traffic flow, we multiply the individual cellular vehicle energy consumption by the number of vehicles in each cell $i$ at time step $t$, yielding:
\begin{equation}
E^{\text{drive}}_{i,t} = e^{\text{drive}}_{i,t} \cdot \rho_{i,t} \cdot \Delta x_i,
\end{equation}
where $\rho_{i,t}$ evolves according to the CTM in \eqref{eq:Si}-\eqref{eq:CTM Update}, capturing spatio-temporal demand under varying traffic conditions. Finally, we include the additional energy required for charging on-board batteries in the EV, resulting in the total energy consumption of the DWC charging coils given by
\begin{equation}
E^{\text{coil}}_{i,t} = E^{\text{drive}}_{i,t} + \left( \mathrm{SoC}^{\text{mile}} \cdot \Delta x_i \cdot E^{\text{battery}} \cdot (\rho_{i,t} \Delta x_i) \right),
\label{eqn:dynamic_energy_consumption}
\end{equation}
where $E^{\text{battery}}$ is the average battery capacity of the EVs in the DWC corridor. To allow additional charging beyond the driving energy, we introduce $\mathrm{SoC}^{\text{mile}}$, representing the extra percentage state-of-charge (SoC) per mile for each vehicle in the roadway. Note that $\mathrm{SoC}^{\text{mile}}$ can be dynamically adjusted based on resource availability, providing flexible energy allocation and real-time responsiveness. This completes our model of
translating non-linear traffic dynamics into charging demand.

\subsection{Microgrid Sizing}\label{subsec: System Sizing}

The microgrid sizing problem can be formulated as the generalized version of the example introduced in Section~\ref{sec: Motivating Example}. Specifically, we formulate a  joint planning and operational optimization problem, where the objective is to determine the optimal capacities of the  solar generation, ES, and external grid coupling of the microgrid that minimizes both capital investment and operating costs. 

Let $\mathcal{N}$ denote the set of buses and $\mathcal{E} \subseteq \mathcal{N}\times\mathcal{N}$ the set of lines in the microgrid. The set $\beta \subseteq \mathcal{N}$ denotes buses where ES installation is allowed. Both the solar generation and the external grid coupling are connected to the slack bus, denoted by $j_0 \!\in\! \mathcal{N}$. This is modeled using an indicator $\delta_{j}$, where $\delta_{j}\! =\! 1$ if $j \!=\! j_0$ and $\delta_{j} \!=\! 0$ otherwise. The operational horizon is $t \!\in\! \mathcal{T} \!=\! \{0,1,\dots,T\}$. The main planning variables are the capacities of the solar generation and grid coupling represented as $P^{\text{S}}$, $P^{\text{G}}$ respectively, and the number of ES units at bus $j$ denoted by $E^{\text{ES}}_j$, with unit costs $C^{\text{S}}$, $C^{\text{G}}$, and $C^{\text{ES}}$ respectively. At each time step $t$, the operational decision variables include nodal real and reactive injections $p_{j,t}, q_{j,t}$, ES charging/discharging powers $p^{\text{ch}}_{j,t}, p^{\text{dis}}_{j,t}$, ES reactive power $q^{\text{ES}}_{j,t}$, ES energy state $e_{j,t}$, external grid injections $p^\text{g}_t, q^\text{g}_t$, and solar active and reactive powers $\gamma_t P^{\text{S}}$ and $ q^\text{s}_t$, where $\gamma_t \! \in \![0,1]$ is the normalized solar availability factor at time step $t$. Then, the power injections at bus $j$ are:
\begin{subequations}\label{eqn:nodal_power_balance}
\begin{align}\label{eq:real_power_balance}
&p_{j,t} \! = \! \delta_{j}(p^\text{g}_{t} \! + \! \gamma_{t} P^{\text{S}}) \!+\! p_{j,t}^\text{dis} \!-\! p_{j,t}^\text{d} \!-\! p_{j,t}^\text{ch} \!+\! \! \! \! \sum_{(j, k) \in \mathcal{E}} \! \! \! \left( p^{\text{flow}}_{j,k,t} \!+\! w_{j,k,t} \right) \!, \\
&q_{j,t} \! = \! \delta_{j}(q^\text{g}_{t} \!+\! q^\text{s}_{t})  \!+\! q_{j,t}^\text{ES}\!-\! q_{j,t}^\text{d} \!+\! \sum_{(j, k) \in \mathcal{E}} \! \! \! \left( q^{\text{flow}}_{j,k,t} \!+\! h_{j,k,t} \right) \!,
\end{align}
$\text{for all } j \in \mathcal{N},\ (j,k) \in \mathcal{E},\ t \in \mathcal{T},$\nonumber
\end{subequations}
where $p^\text{d}_{j,t}$ and $q^\text{d}_{j,t}$ denote the real and reactive power demand from the DWC at bus $j$, and $w_{j,k,t} := r_{j,k} \ell_{j,k,t}$ and $h_{j,k,t} := x_{j,k} \ell_{j,k,t}$ represent the real and reactive power losses of line $(j,k) \in \mathcal{E}$, where $r_{j,k},x_{j,k}$ are its respective resistance and reactance, and $\ell_{j,k,t}$ is the squared current magnitude at time $t$. 
For each $t \in \mathcal{T}$, let $\mathbf{p}_t, \mathbf{q}_t \!\in\! \mathbb{R}^{|\mathcal{N}|}$ be the vectors obtained by stacking the real and reactive injections $p_{j,t}$ and $q_{j,t}$ at buses $j \in \mathcal{N}$, respectively. Then, we  say that $(\mathbf{p}_t,\mathbf{q}_t) \!\in \!\mathcal{S}_t$, where  $\mathcal{S}_t$ is termed the set of \textit{feasible} real and reactive power injection vectors, iff there exist nodal power injections $p_{j,t}, q_{j,t}$, squared voltage magnitudes $v_{j,t}, \forall \! j \! \in \! \mathcal{N}, \forall t \! \in \! \mathcal{T}$, branch flows $p^{\text{flow}}_{j,k,t}, q^{\text{flow}}_{j,k,t}$, and squared current magnitudes $\ell_{j,k,t}$ for all lines $(j,k) \! \in\!  \mathcal{E}$ satisfying \eqref{eqn:nodal_power_balance} such that the following operational constraints hold for all $\text{for all } j \in \mathcal{N},\ (j,k) \in \mathcal{E},\ t \in \mathcal{T}$: 
\begin{subequations}\label{eq:voltage_line_constraints}
\begin{align}\label{eq:voltage_drop_new}
&v_{k,t} = v_{j,t} - v^{\text{drop}}_{j,k,t} + \big( (r_{j,k})^2 + (x_{j,k})^2 \big) \ell_{j,k,t}, \\[1ex]\label{eqn:socp_relaxation}
&\left\| 
\begin{bmatrix}
2 p^{\text{flow}}_{j,k,t} & 
2 q^{\text{flow}}_{j,k,t} & 
(\ell_{j,k,t} - v_{j,t})
\end{bmatrix}^{\mathsf{T}}
\right\|_2
\leq \ell_{j,k,t} + v_{j,t}, \\[1ex]\label{eqn:voltage_limits}
&\underline{v}_j \leq v_{j,t} \leq \overline{v}_j, \quad \ell_{j,k,t} \leq \overline{\ell}^{j,k},\\[1ex]
&0 \!\leq\! p^{\text{g}}_{t} \!\leq\! P^\text{G},  \, - \overline{q}^{\text{g}} \!\leq\! q^{\text{g}}_{t} \!\leq\! \overline{q}^{\text{g}}, \,
- \overline{q}^{\text{s}} \!\leq\! q^{\text{s}}_t \!\leq\! \overline{q}^{\text{s}}, \label{eq:power_bounds}
\end{align}
\end{subequations}
where ~\eqref{eq:voltage_drop_new} characterizes the line voltage drop, with 
$v^{\text{drop}}_{j,k,t} := 2\,(p^{\text{flow}}_{j,k,t} r_{j,k} + q^{\text{flow}}_{j,k,t} x_{j,k})$ denoting the linear contribution from real and reactive power flows, 
and $\big((r_{j,k})^2 + (x_{j,k})^2\big)\ell_{j,k,t}$ capturing the quadratic loss along line $(j,k)\in\mathcal{E}$. 
Equation~\eqref{eqn:socp_relaxation} provides an SOCP relaxation of the nonconvex voltage-current relationship in AC power flow for tractability~\cite{farivarBranchFlowModel2013}, 
while \eqref{eqn:voltage_limits} enforces safe operating limits and \eqref{eq:power_bounds} ensures that the  active and reactive powers respect their corresponding limits.

Similarly, for the buses equipped with ES, denoted by $j \!\in\! \beta$, we define 
$\mathbf{p}^{\text{ch}}_t, \mathbf{p}^{\text{dis}}_t, \mathbf{q}^{\text{ES}}_t \in \mathbb{R}^{|\beta|}$ 
as the stacked vectors of charging, discharging, and reactive powers $p^{\text{ch}}_{j,t}, \, p^{\text{dis}}_{j,t}$ and $q^{\text{ES}}_{j,t}$ respectively. For buses $k \notin \beta$, the corresponding ES variables are zero for all  $ t \in \mathcal{T}$. Then, we say that  
$(\mathbf{p}^{\text{ch}}_t, \mathbf{p}^{\text{dis}}_t, \mathbf{q}^{\text{ES}}_t) \in \mathcal{F}_t$, where $\mathcal{F}_t$ is termed the feasible set of ES charging, discharging, and reactive powers at time step $t$ iff:
\begin{subequations}\label{eq:battery_constraints}
\begin{align}
&e_{j,t} = e_{j,(t-1)} + \Delta t \left( \eta^{\text{ch}} p^{\text{ch}}_{j,t} - \frac{1}{\eta^{\text{dis}}} p^{\text{dis}}_{j,t} \right), \; \forall t \in \mathcal T \backslash \{T\}, \label{eq:energy_soc_update} \\
&e_{j,0} = e_{j,T}, \; \;0 \leq e_{j,t} \leq \overline{e}^{\text{unit}} \!\cdot\! E^{\text{ES}}_j,\label{eq:energy_dynamics_combined} \\[2ex]
&0 \leq p^{\text{ch}}_{j,t} \leq \overline{p}^{\text{ch}}_j, \; 0 \leq p^{\text{dis}}_{j,t} \leq \overline{p}^{\text{dis}}_j, \;
-\overline{q}^{\text{ES}}_{j} \leq q^{\text{ES}}_{j,t} \leq \overline{q}^{\text{ES}}_{j},\label{eq:es_energy_constraint}
\end{align}
$\text{for all } j \in \beta,\ t \in \mathcal{T},  E^{\text{ES}}_j \in \mathbb{Z}_{\geq 0}.$
\end{subequations}

Here, equation~\eqref{eq:energy_soc_update} describes the SoC evolution of the ES at bus $j$ at each time step $t$, where $\eta^{\text{ch}}$ and $\eta^{\text{dis}}$ denote the charging and discharging efficiencies, respectively. Equation~\eqref{eq:energy_dynamics_combined} enforces equal initial and terminal SoC over the operational horizon and constrains the SoC within the installed energy capacity, thereby ensuring periodic operation and feasible cycling. The parameter $\overline{e}^{\text{unit}}$ represents the rated capacity of a candidate ES unit, and $E^{\text{ES}}_j \in \mathbb{Z}_{\geq 0}$ denotes the integer number of ES units installed at bus $j \in \beta$. In practice, this variable is often relaxed to continuous values to improve convexity and computational tractability. Finally, \eqref{eq:es_energy_constraint} limits the charging, discharging, and reactive power outputs of the ES within their respective rated bounds.

The external grid is modeled as a “virtual” generator with cost function $C_g(p^{\text{g}}_t) := a (p^{\text{g}}_{t})^2 + b p^{\text{g}}_{t} + c$, where $a$, $b$, and $c$ are its quadratic cost coefficients. Since solar generation and ES are renewable, they incur no direct operating cost; however, a soft penalty $\lambda$ is introduced on ES charging and discharging to discourage infeasible behavior such as simultaneous operation. Then, the operational problem is then formulated as an AC Optimal Power Flow (AC-OPF) as follows,  where the objective is to minimize operational costs in the microgrid, while ensuring that the grid constraints are satisfied at all times.
\begin{subequations}\label{eq:operational_problem}
\renewcommand{\theequation}{\theparentequation\alph{equation}}
\begin{align}
\mathcal{J} = 
\min &\sum_{t \in \mathcal{T}} \Biggl[ C_g(p^{\text{g}}_t) + \lambda \sum_{j \in \beta} \left( p^{\text{ch}}_{j,t} + p^{\text{dis}}_{j,t} \right) \Biggr],\label{eq:objective_function}\\[4pt]
\text{s.t.} \quad
&(\mathbf{p}_t,\mathbf{q}_t) \in \mathcal{S}_t, 
\quad \quad \quad \quad \forall t \in \mathcal{T} \\
&(\mathbf{p}^{\text{ch}}_t, \mathbf{p}^{\text{dis}}_t, \mathbf{q}^{\text{ES}}_t) \in \mathcal{F}_t. \quad \, \forall t \in \mathcal{T}
\end{align}
\end{subequations}
In \eqref{eq:operational_problem}, $\mathcal{J}$ denotes the optimal operating cost of the microgrid over the time horizon $\mathcal{T}$, subject to the grid constraints captured by the sets $\mathcal{S}$ and $\mathcal{F}$. This operating cost is then used to formulate the microgrid planning problem, analogous to \eqref{eq:motivating_optimization}, by incorporating $\mathcal{J}$ over an operational lifetime of the infrastructure, denoted by  $\mathcal{P}$, typically expressed in the number of years. The microgrid planning problem is written as
\begin{equation}\label{eq:planning_problem}
\begin{aligned}
\min_{P^{\text{S}}, P^{\text{G}}, \{E^{\text{ES}}_j, \, j \in \beta\}} \quad
& C^{\text{S}} P^{\text{S}} 
+ C^{\text{ES}} \sum_{j \in \beta} E^{\text{ES}}_j 
+ C^{\text{G}} P^{\text{G}} 
+ \mathcal{P} \cdot \mathcal{J}, \\
\text{s.t.} \quad
& P^{\text{S}}, P^{\text{G}} \ge 0, \quad
E^{\text{ES}}_j \in \mathbb{Z}_{\geq 0}, \; \forall j \in \beta.
\end{aligned}
\end{equation}
The problem~\eqref{eq:planning_problem} yields the optimal nonnegative capacities of microgrid solar generation, ES, and external grid coupling, thereby completing the sizing formulation.

\subsection{Microgrid Validation}\label{subsec: Microgrid Validation}

After solving the microgrid planning problem \eqref{eq:planning_problem}, it is necessary to validate the feasibility and robustness of the obtained design under different demand scenarios. This is carried out by solving the operational problem for each scenario and verifying that all network constraints are satisfied over the entire time horizon. For this validation,  we introduce slack variables  that capture any mismatch between supply and demand when the system is unable to strictly satisfy the power balance constraints in \eqref{eqn:nodal_power_balance} due to infeasibility under a given demand scenario. The slack variables act as a diagnostic measure of infeasibility, allowing the identification and quantification of unmet demand while the planned system remains unchanged. A threshold $\alpha$ is then imposed on the aggregate slack, accounting for solver precision, and scenarios with total slack below this threshold are considered successfully validated.

Let $s^{\text{p}}_{j,t}$ and $s^{\text{q}}_{j,t}$ denote the slack variables corresponding to the real and reactive power injections in bus $j$ at time step $t$, respectively. These variables are then incorporated into the original nodal real and reactive power injections from \eqref{eqn:nodal_power_balance} as:
\begin{subequations}\label{eqn:nodal_power_balance_val}
\begin{align}\label{eq:real_power_balance_val}
&p^{\text{val}}_{j,t}  =  \delta_{j}(p^\text{g}_{t} + p^{\text{s}}_t) \! + p_{j,t}^\text{dis} - p_{j,t}^\text{d} - p_{j,t}^\text{ch} \nonumber \\
&\quad \quad \quad \quad \quad \quad \quad \quad \, \,  + \sum_{(j, k) \in \mathcal{E}} \! \left( p^{\text{flow}}_{j,k,t} + w_{j,k,t} \right) + s^{\text{p}}_{j,t}, \\
&q^{\text{val}}_{j,t}  =  \delta_{j}(q^\text{g}_{t} + q^\text{s}_{t}) +q_{j,t}^\text{ES} - q_{j,t}^\text{d}\nonumber\\
& \quad \quad \quad \quad \quad \quad \quad \quad \, \,    + \sum_{(j, k) \in \mathcal{E}}  \! \left( q^{\text{flow}}_{j,k,t} + h_{j,k,t} \right) + s^{\text{q}}_{j,t},
\end{align}
$\text{for all } j \in \mathcal{N},\ (j,k) \in \mathcal{E},\ t \in \mathcal{T}.$\nonumber
\end{subequations}

Here, $p^{\text{s}}_t \!\leq\! \gamma_t P^{\text{S}}$ represents the controlled solar injection constrained by the total available generation at time $t$. Then, we redefine the set of feasible real and reactive injection vectors, denoted as $\mathbf{p}_t^\textbf{val}, \mathbf{q}_t^\textbf{val} \in \mathbb{R}^{|\mathcal{N}|}$ satisfying the modified power balance equations in \eqref{eqn:nodal_power_balance_val}.  We then minimize the total system slack by modifying the objective  of the operational problem in \eqref{eq:objective_function} as follows:
\begin{subequations}\label{eq:operational_problem_val}
\renewcommand{\theequation}{\theparentequation\alph{equation}}
\begin{align}
\min &\sum_{t \in \mathcal{T}} \Biggl[ C_g(p^{\text{g}}_t) \!+\! \lambda \!\sum_{j \in \beta}\! \left( p^{\text{ch}}_{j,t} \!+\! p^{\text{dis}}_{j,t} \right)\! +\! \Theta \! \sum_{j\in \mathcal{N}} \!\left(s^{\text{p}}_{j,t}\! + \!s^{\text{q}}_{j,t} \right)\!\Biggr],\\[4pt]
\label{eq:objective_function_val}
\text{s.t.} \quad
&(\mathbf{p}_t^\textbf{val},\mathbf{q}_t^\textbf{val}) \in \mathcal{S}_t, 
\quad \quad \, \, \, \, \forall t \in \mathcal{T} \\
&(\mathbf{p}^{\text{ch}}_t, \mathbf{p}^{\text{dis}}_t, \mathbf{q}^{\text{ES}}_t) \in \mathcal{F}_t, \quad \forall t \in \mathcal{T}
\end{align}
\end{subequations}
where, $\Theta$ denotes a penalty on the  the  slack. After running the validation problem formulated in \eqref{eq:operational_problem_val}, we impose a threshold $\alpha$ on the total slack for each demand scenario, and say that  a given demand scenario is considered valid iff the total slack remains below this threshold, that is, 
\begin{equation}\sum_{t \in \mathcal{T}} \sum_{j \in \mathcal{N}} s^{\text{p}}_{j,t} \leq \alpha, \text { and }\sum_{t \in \mathcal{T}} \sum_{j \in \mathcal{N}} s^{\text{q}}_{j,t} \leq \alpha.\end{equation}
All other scenarios are classified as failed, thereby completing the microgrid validation process.

\begin{figure}[t]
    \centering
    \includegraphics[width=0.5\textwidth]{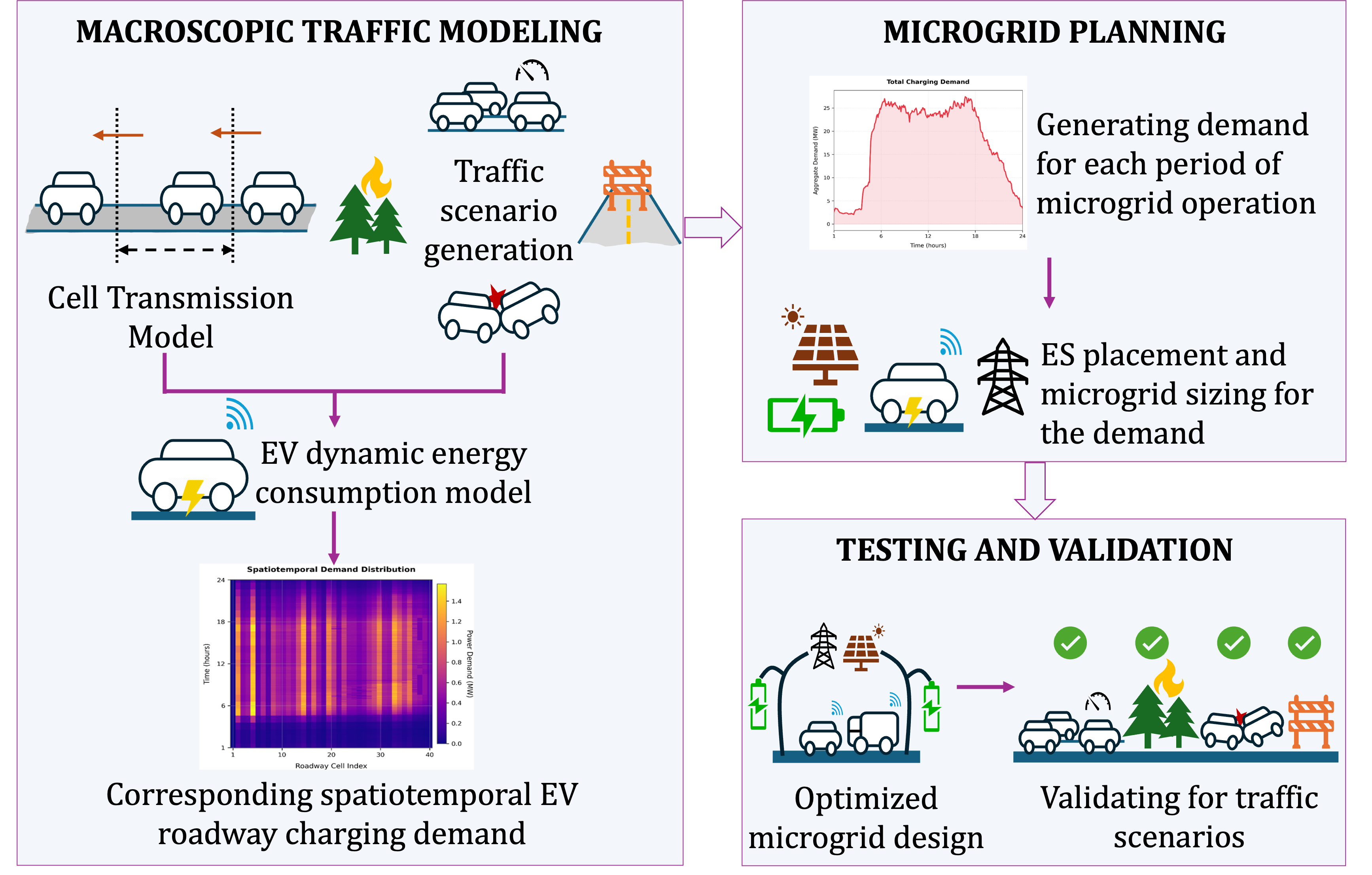}
    \caption{Proposed methodological framework for traffic-aware microgrid planning of a DWC system. The approach employs a macroscopic CTM to represent traffic flow dynamics, which is then used to derive the spatiotemporal energy demand along the roadway. These demand profiles serve as inputs for microgrid planning and parameterization, and are subsequently evaluated and validated under multiple traffic scenarios to assess system robustness.}
    \label{fig:methodology}
    \vspace{-0.5cm}
\end{figure}

Fig.~\ref{fig:methodology} illustrates the overall planning and validation framework, summarizing the developments in Sections~\ref{subsec: Cell Transmission Model}--\ref{subsec: Microgrid Validation}.

\section{Case Study - Set-Up}\label{sec: Case Study - Set-Up}

As a case study for the proposed framework, we consider a 14-mile segment of Interstate 210 West (I-210W) in California, extending from Vernon to the junction where I-210 diverges from State Route 134 (SR-134). This corridor is a major highway in the densely populated Los Angeles region and is therefore a good representative corridor for evaluating DWC deployment on U.S. highways. The segment is also well monitored through the Performance Measurement System (PeMS) operated by Caltrans, which provides regular traffic flow data~\cite{CaltransPeMS}, and has pre-calibrated CTM parameters available through the CTMSIM package~\cite{CTMSIMInteractiveFreeway2025}. Since CTM calibration is a data-intensive process beyond the scope of this work, leveraging data from this existing well-calibrated and validated corridor allows us to obtain realistic results in this case study. 

\subsection{System Design}\label{subsec: System Design}

\begin{figure}[!t]
    \centering
    \includegraphics[width=0.50\textwidth]{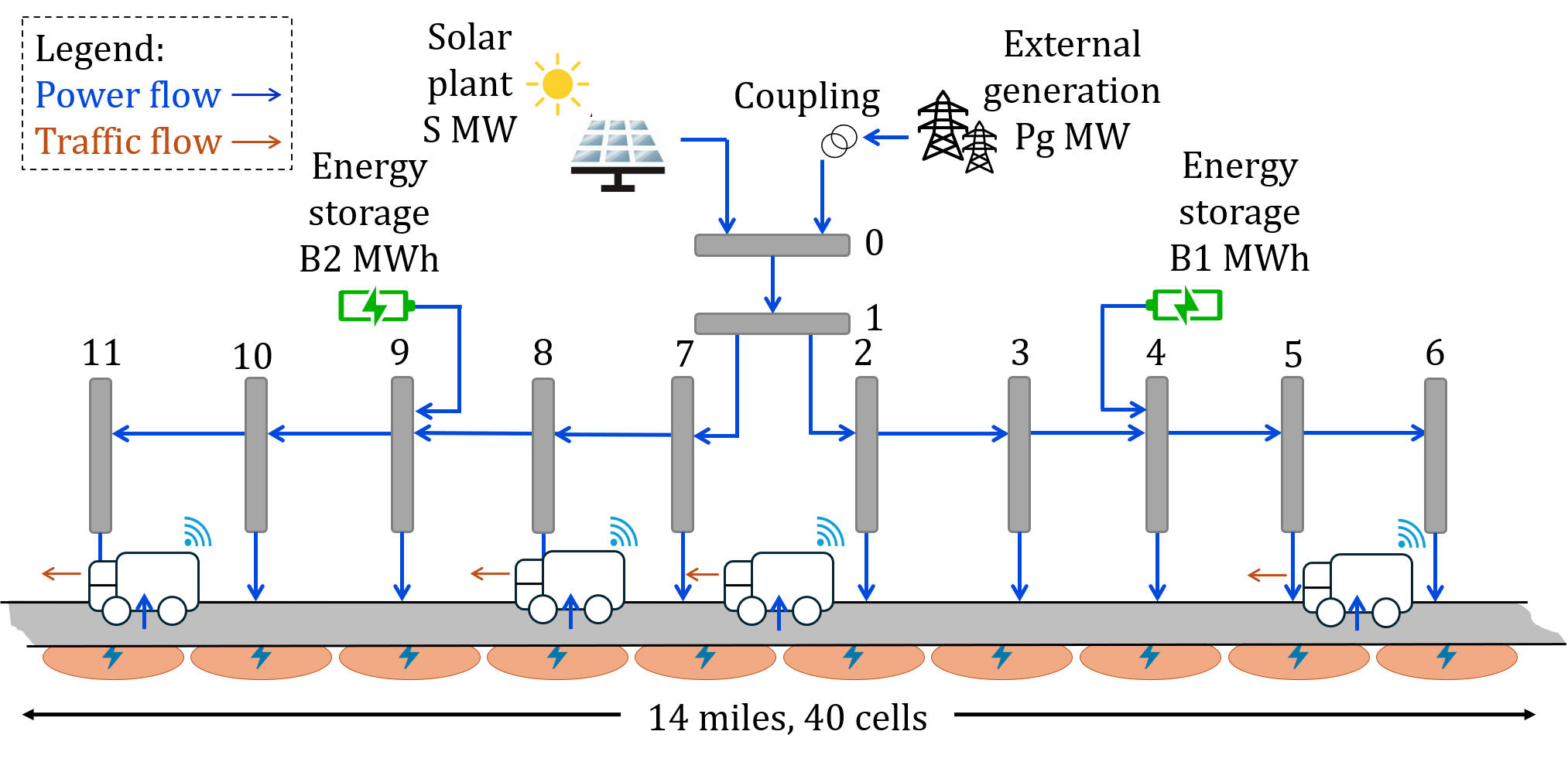}
    \caption{The proposed 12-bus system architecture for the case study. The roadway includes a single charging lane of length 14 miles, divided into 40 cells, with each bus parallel to the roadway serving 4 cells. The slack bus (bus 0) is connected to the solar generation and the coupled external grid.}
    \label{fig:radialgrid}
    \vspace{-0.5cm}
\end{figure}

The 14-mile I-210 W segment is divided into 40 cells, with lengths determined using CTM calibration and historical traffic flow values~\cite{CTMSIMInteractiveFreeway2025}. The supporting microgrid is modeled as a 12-bus radial network as shown in Fig.~\ref{fig:radialgrid}. The Buses 2–11 are aligned along the roadway, each supplying DWC loads associated with four traffic cells, while Buses 0 and 1 serve as the interface to the external grid and solar generation. Central Buses 0 and 1 are placed 5 miles apart, and two branches extend from Bus 1 to Buses 2 and 7, each two miles away. All remaining buses are connected sequentially with 1.4-mile spacing between adjacent nodes. Transmission line resistances and reactances scale with the bus distances. ES units will be placed on the remaining buses by solving the ES siting optimization problem, with placements in Fig. \ref{fig:radialgrid} shown for illustration only. Since the focus is on capacity planning rather than exact spatial correspondence, a representative mapping between traffic cells and buses is sufficient to capture grid capacity, dispatch behavior, and infrastructure sizing. The planning problem as defined in \eqref{eq:planning_problem} is evaluated over a 20-year planning horizon.

\subsection{Traffic Scenarios}\label{subsec: Traffic Scenarios}

\begin{figure}[t]
\centering
\includegraphics[width=0.49\textwidth]{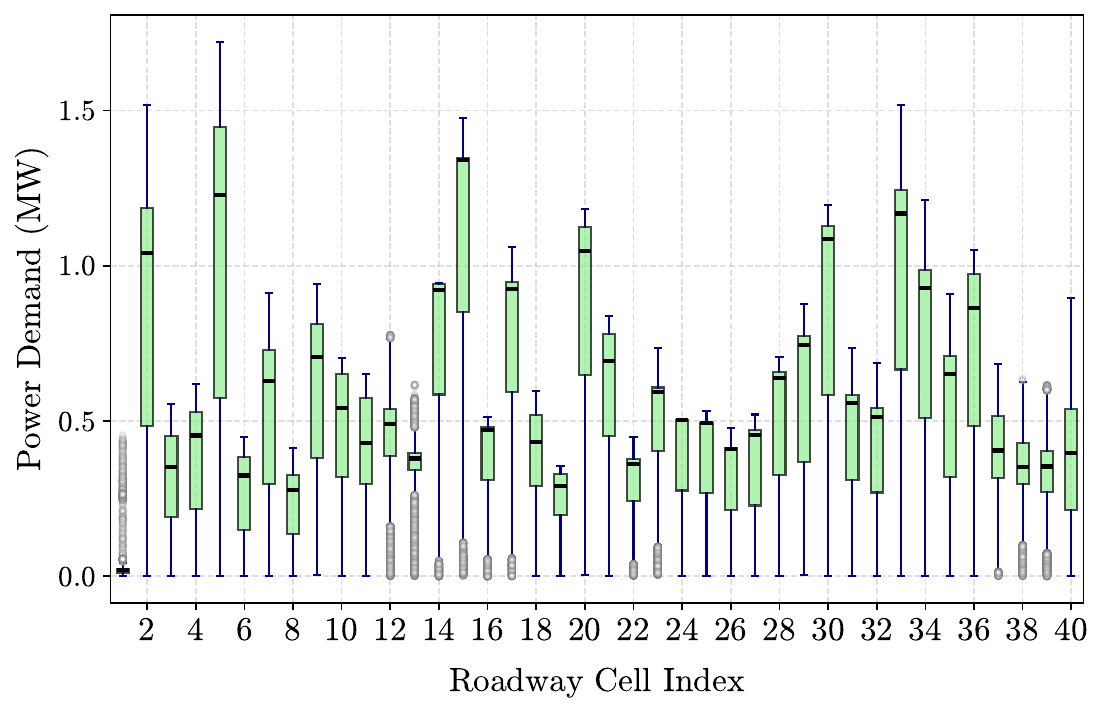}
\caption{Distribution of the charging demand in each cell across the 100 scenarios. The differences in the distribution patterns across cells highlight the spatiotemporal characteristics of traffic flow. In the figure, the box represents the 25th to 75th percentile range, the central lines in the boxes indicate the median, and the whiskers denote the minimum and maximum demand values.}
\label{fig:energy_variations}
\vspace{-0.5cm}
\end{figure}

To assess the reliability of the DWC microgrid, we simulate traffic scenarios that induce fluctuations in charging demand. Scenario formulation is non-trivial and involves adjusting the CTMSIM parameters to capture traffic fluctuations. Although traffic is sensitive to localized events and real-time driver responses, making purely analytical predictions difficult, we construct representative scenarios that capture the essential traffic dynamics and provide a meaningful basis for simulation and evaluation. Based on documented events in this region and typical highway behavior, we classify these scenarios into four categories, representing distinct conditions or disruptions:
\begin{enumerate}[label=\alph*)]
\item \textbf{Scenario I:} Representative of regular day-long demand fluctuations simulated by scaling and adding Gaussian noise to ramp inflows and outflows, initial densities, and demand profiles.
\item \textbf{Scenario II:} Representative of planned, all-day disruptions in certain cells modeled by reducing on- and off-ramp capacities in affected areas, such as scheduled lane blockages or closures.
\item \textbf{Scenario III:} Representative of unplanned disruptions of varying time, such as accidents. Mainline disruptions can be Mild, Moderate, or Severe, with probabilities of 30\%, 30\%, and 40\%, respectively, defined by the number of lanes blocked. A small fraction of scenarios affects only ramps, creating short, localized fluctuations in ramp flow.
\item \textbf{Scenario IV:} Representative of high-impact events, such as evacuations, simulated by blocking the West end of the roadway and varying the incoming traffic.
\end{enumerate}
Fig.~\ref{fig:energy_variations} illustrates the impact of traffic variability on power demand that the microgrid must accommodate, where the large fluctuations correspond to recurring disruptions or cells near ramps. We note that no clear trend is observed across the corridor, replicating the dynamic nature of real-world traffic.

\subsection{Dataset Description}\label{subsec: Dataset Description}
This case study focuses on Class 9 heavy-duty EVs (HEVs), with vehicle parameters from~\cite{haddad2022, IEA2023}. The DWC system is built on the rightmost lane, with HEVs comprising 12\% of traffic~\cite{StatesHighestLowest} and 90\% of them using the charging lane~\cite{californiaTruckLaneUse}. Traffic data for the segment is obtained from the CTMSIM toolbox~\cite{CTMSIMInteractiveFreeway2025}, and scaled for different scenarios. Solar data uses normalized output from an 80 MW California plant~\cite{SolarPowerData}. Microgrid operational and capital costs are derived from the Texas A\&M synthetic grid~\cite{ACTIVSg10k10000busSynthetic}, NREL 2023 solar and ES benchmarks~\cite{ramasamyUSSolarPhotovoltaic2023, MegapackUtilityScaleEnergy, PNNL2025LiIonBattery}, and the 2018 NREL microgrid planning study~\cite{giraldezminerPhaseMicrogridCost2018}, all scaled for inflation. The line resistance and reactance are computed from per-mile Osprey ACSR conductor parameters~\cite{graingerPowerSystemAnalysis1994} weighted by inter-bus distances.
\section{Case Study - Results}\label{sec: Case Study - Results}
We now solve the microgrid planning problem and validate our design across different traffic scenarios. 

\subsection{Dynamic Energy Consumption}\label{subsec: Dynamic Energy Consumption}

\begin{figure}[t]
    \centering
    \includegraphics[width=\linewidth]{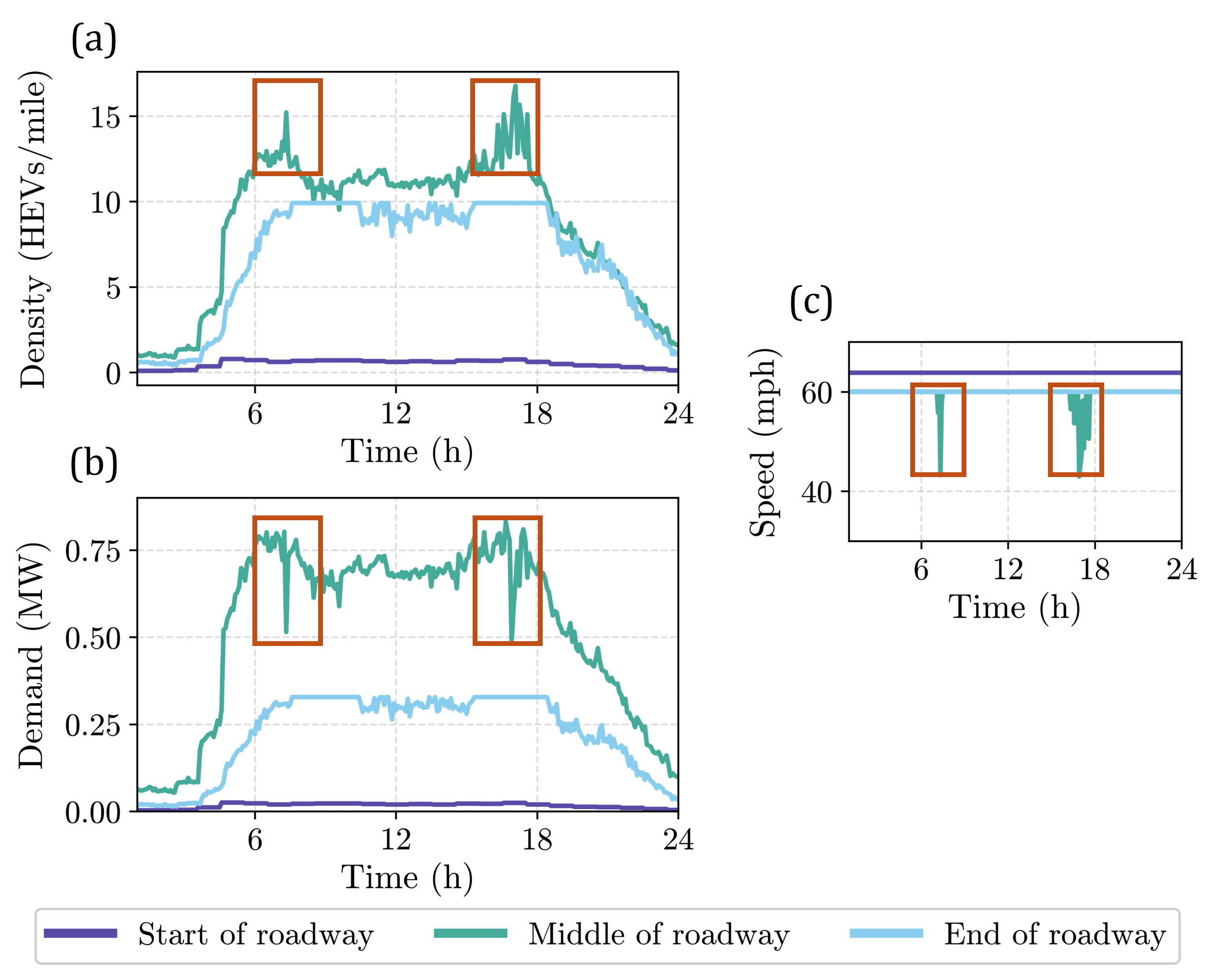}
    \caption{Plots demonstrating the nonlinear relationship between traffic density and the corresponding DWC demand: (a) shows the density in representative roadway cells, (b) shows the corresponding demand, and (c) shows the average vehicle speed in those cells. The insets illustrate that as traffic density increases in (a), vehicle speeds begin to decrease beyond a threshold in (c), leading to a reduction in demand in (b).}
    \label{fig:traffic_density}
    \vspace{-0.5cm}
\end{figure}

Fig. \ref{fig:traffic_density} illustrates the non-linear relationship between traffic density and aggregate charging demand. While energy demand generally follows traffic density under normal conditions, this relationship weakens during congestion due to the model in Section~\ref{subsec: Dynamic Energy Consumption Model}, where 
\( p_{\text{drive}} \) depends on the cube of the traffic density and velocity as in \eqref{eqn: dynamic energy consumption}. For this simulation, we use baseline CTMSIM data and examine representative cells from the beginning, middle, and end of the highway over a 24-hour horizon. In general, a higher vehicle density, as shown in the density-time plot in Fig. \ref{fig:traffic_density}, results in higher total power demand since more vehicles draw energy from DWC. However, deviations occur when density spikes coincide with reductions in power demand, two of which are highlighted in the inset of Fig. \ref{fig:traffic_density}. These arise during congestion when vehicle speeds drop sharply. Because per-vehicle power demand depends on both density and velocity, lower speeds reduce total energy consumption, as defined in \eqref{eqn: dynamic energy consumption}. The average speed-time plot in Fig. \ref{fig:traffic_density} confirms this behavior, showing pronounced declines in velocity at the same times as the density peaks. For this roadway segment, the maximum demand reaches approximately 27.8 MW. The nonlinear relationship between traffic flow and energy consumption illustrated here underlines the need for accurate traffic-aware demand modeling  and forms the basis for the following microgrid design.

\subsection{Microgrid Planning}\label{subsec: Grid Modeling}

The microgrid planning task can be divided into two stages: the placement of ES units and the sizing of the system based on the identified buses equipped with ES units.

\subsubsection{ES Placement}\label{subsec: Battery Placement}

The goal of the ES placement problem is to identify the subset of buses $\beta \!\subseteq\! \mathcal{N}$ where reactive power support is most critical, thereby selecting appropriate nodes for ES placement. This placement is important because ES units serve a dual purpose: they provide a flexible power source to meet the spatiotemporal demand from the DWC and supply reactive power at each bus according to traffic-driven consumption patterns, supporting voltage stability. Proper placement is therefore essential, as the spatial distribution of ES units affects distances between components, which in turn influences line losses and overall system efficiency. As a result, the placement should align with traffic behavior, locating ES units at points where voltage drops are more likely to occur.

\begin{figure}[t]
    \centering
    \includegraphics[width=\linewidth]{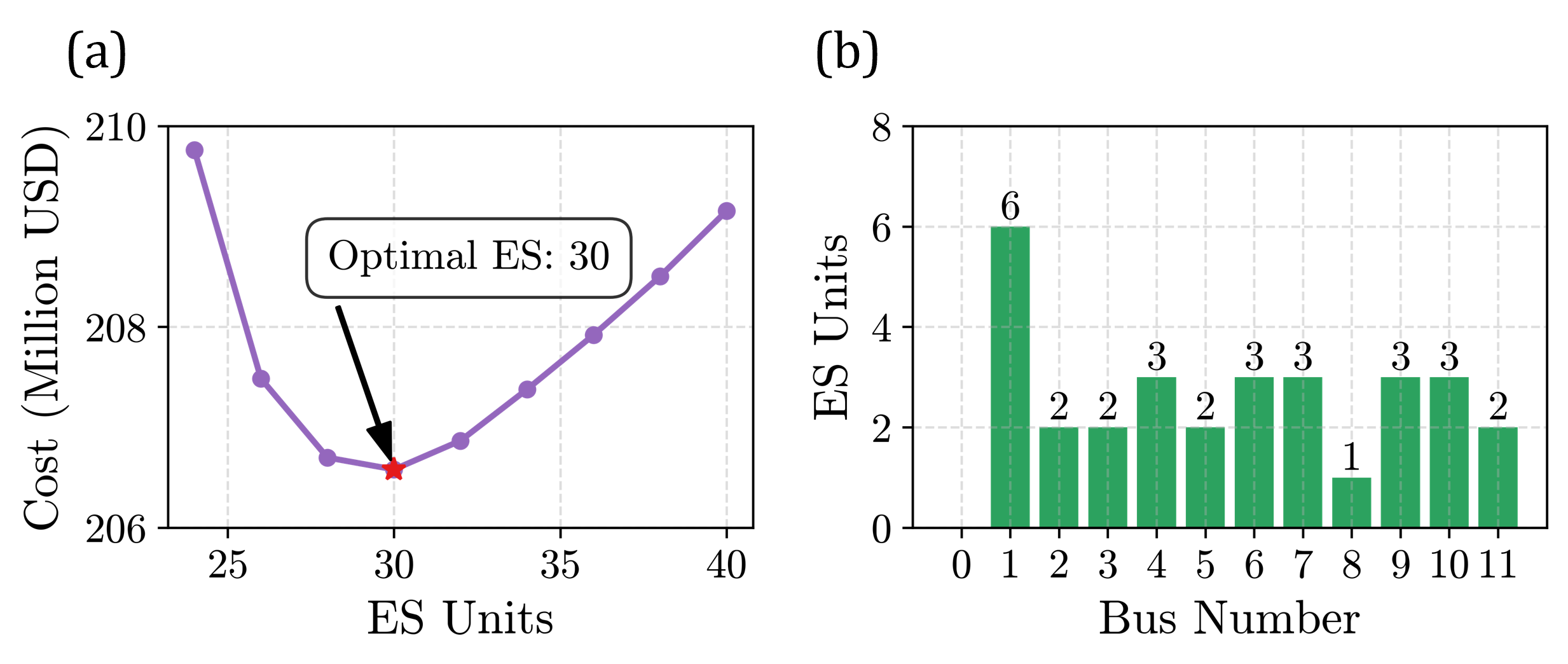}
    \caption{Results for the ES placement optimization: (a) shows the variation of system cost with the total number of ES units $k_{\text{batt}}$, and (b) shows the distribution of ES units for the optimal $k^{\text{ES}} = 30$. The selected buses form the set $\beta \subseteq \mathcal{N}$ used in the planning problem.}
    \label{fig:placement_cost_curve}
    \vspace{-0.5cm}
\end{figure}

To achieve this, the optimization problem defined in \eqref{eq:planning_problem} is solved by varying the total number of ES units $k^{\text{ES}} = \sum_{j\in\mathcal{N}} E^{\text{ES}}_j$ allocated to the system until an optimal number of ES units with the corresponding minimum cost is identified. For the DWC load, one of the baseline CTMSIM datasets~\cite{CTMSIMInteractiveFreeway2025}, which contains calibrated traffic data from PeMS on I-210W, is used. The optimal configuration consists of 30 units, as shown in Fig.~\ref{fig:placement_cost_curve}(a), which minimizes the overall cost. The results reflect the trade-off between the operational flexibility provided by additional ES units and their installation costs.

Next, we examine the distribution of the 30 ES units to identify preferred buses for ES installation in this representative case. The resulting allocation is shown in Fig.~\ref{fig:placement_cost_curve}(b). Given the radial structure in Fig.~\ref{fig:radialgrid}, Bus 0 is excluded from installation because its larger electrical distance from the remaining buses increases transmission losses and total cost. Bus 1, which is more centrally located, receives the largest share of ES units. The subsequent units are distributed nearly uniformly among buses serving the highway cells, with minor differences driven by the local DWC load. For the remaining study, the selected ES buses in Fig.~\ref{fig:placement_cost_curve}(b) are fixed.

\subsubsection{Microgrid Parameters for Baseline Traffic Flows}\label{subsubsec: Default PeMS-based Traffic System}
\begin{figure}[t]
    \centering
    \includegraphics[width=\linewidth]{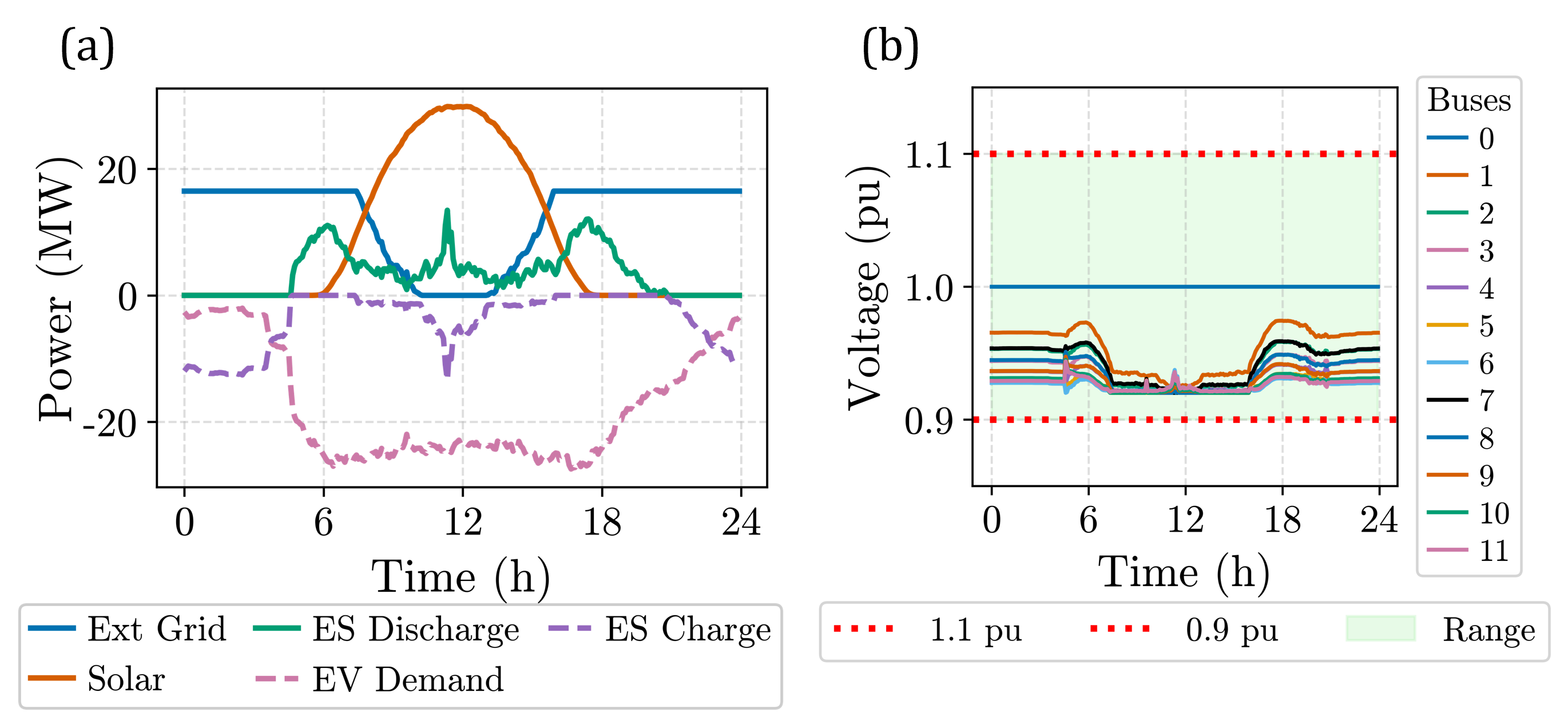}
    \caption{Baseline traffic simulation results showing microgrid power generation and bus voltage profiles: (a) illustrates the microgrid power balance, with peak solar injection during daytime and increased ES discharge in the evening hours; (b) shows the variation in voltage profiles across different buses due to DWC demand. All voltages remain within safe operating limits..}
    \label{fig:base_case_parameters}
    \vspace{-0.5cm}
\end{figure}

Fig.~\ref{fig:base_case_parameters}(a) presents the sizing results based on the baseline CTMSIM traffic data. The optimal design includes 43.41 MW of solar capacity and 117 MWh of total ES capacity. An external grid connection of about 16.36 MW is required to maintain reliability during peak demand. The microgrid power balance curve illustrates the interaction among solar generation, ES, and grid imports. Solar output peaks around midday, reducing grid withdrawals to nearly zero. Excess solar energy charges the ES units, which discharge during the evening peak as solar generation declines. Grid imports meet residual demand when solar generation and ES are insufficient. The charging and discharging values shown are aggregated across all ES units.
Fig.~\ref{fig:base_case_parameters}(b) depicts the voltage profiles for all buses in the system over the operational period. All voltages remain within acceptable limits, demonstrating that the system is effectively designed to accommodate spatiotemporal demand fluctuations. Minor voltage drops are observed during high demand periods, which are mitigated by the reactive power supplied from the solar inverters and the strategic placement of ES units.

\subsubsection{Traffic Scenario Simulations}\label{subsubsec: Traffic Scenario Simulations}

\begin{figure}[t]
    \centering
    \includegraphics[width=0.48\textwidth]{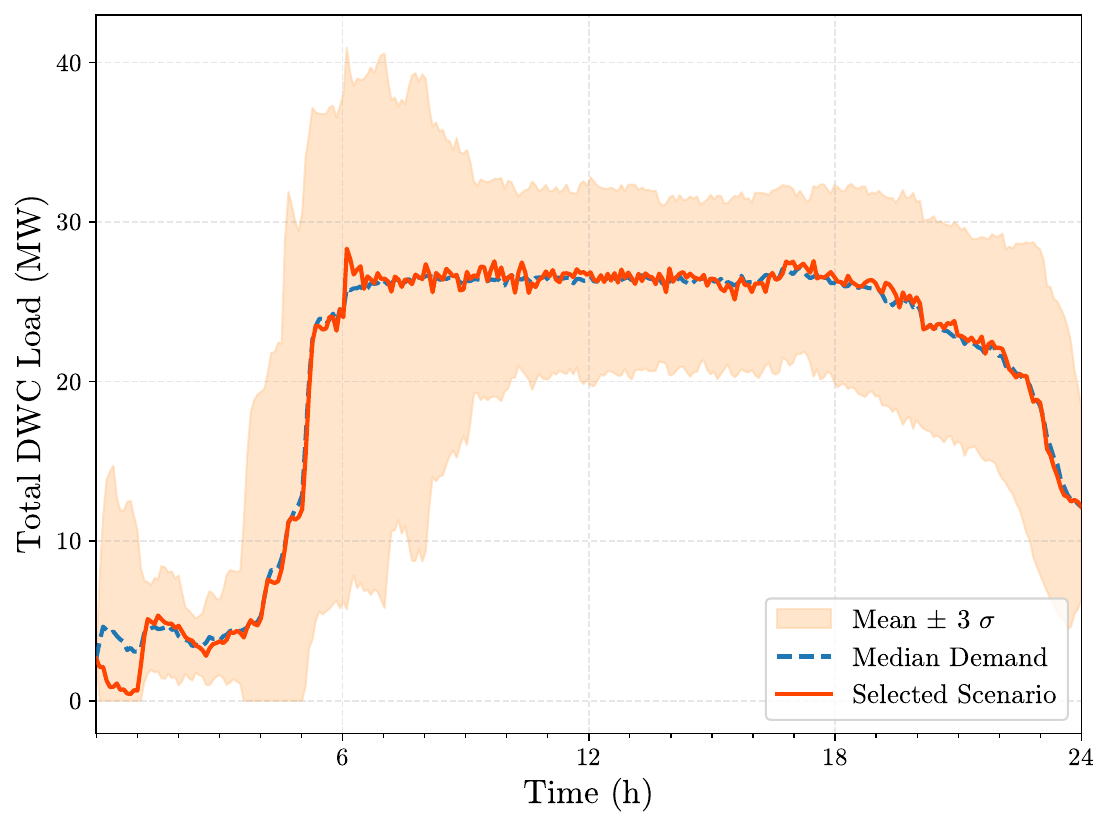}
    \caption{Variation of the total DWC load across all nodes at each time step. The dotted blue line represents the median total demand across all time steps, while the red line indicates the selected planning scenario with total load closest to this median. This selected scenario is used to capture the corresponding spatiotemporal distribution of load across all cells and time steps.}
    \label{fig:DWC_load_dist}
    \vspace{-0.5cm}
\end{figure}

 Fig.~\ref{fig:DWC_load_dist} plots the distribution of the DWC demand profiles for the traffic scenarios defined in Section~\ref{subsec: Traffic Scenarios}. A total of 100 different scenarios are evaluated: Type I (30 runs), Type II (20 runs), Type III (35 runs), and Type IV (15 runs). Each scenario affects traffic flow differently, leading to distinct energy demand profiles. This distribution is used to identify the median total DWC demand in the corridor. The variation in total demand, shown in Fig.~\ref{fig:DWC_load_dist}, further highlights the variability in traffic and motivates the need to couple traffic modeling with power system operation for DWC infrastructure planning. The median total demand is then compared across all 100 scenarios to identify a representative scenario that most closely matches this profile, allowing us to obtain the representative spatiotemporal distribution of demand at each node and time step. 
The demand from this representative scenario is then used to solve the planning problem.

\subsection{Final Microgrid Design and Validation}\label{subsec: Validating Selected System's Performance Against Varying Traffic Scenarios}
Having identified the representative DWC demand profile for the microgrid, we use the planning problem defined in \eqref{eq:planning_problem} to determine the proposed microgrid parameters. Solving the AC-OPF over a planning horizon of 20 years yields a required solar installation of 47.5 MW, a total ES installed capacity of 132.3 MWh, and 21 MW of external grid injection. The distribution of ES across different buses is shown in Table~\ref{tab:es_capacity}, where Bus 0 has no ES units, Bus 1 receives the largest share, and the remaining buses have approximately equal capacities. The ES variables are modeled as continuous rather than integer variables for computational tractability. While a more detailed formulation could refine sizing, over a 20-year horizon, small variations have limited impact on the overall installed capacity.

\begin{table*}
\centering
\caption{Final distribution of ES capacities across the different buses in the proposed system from Fig.~\ref{fig:radialgrid}}
\label{tab:es_capacity}
\setlength{\tabcolsep}{10pt}
\small 
\begin{tabular}{@{}lcccccccccccc@{}}
\toprule
\textbf{Bus Index} & 0 & 1 & 2 & 3 & 4 & 5 & 6 & 7 & 8 & 9 & 10 & 11 \\ \midrule
\textbf{Capacity (MWh)} & 0 & 30.9 & 11.4 & 12.15 & 8.8 & 10.7 & 8.0 & 9.3 & 8.7 & 10.4 & 14.8 & 7.2 \\ \bottomrule
\end{tabular}
\end{table*}

\begin{figure}[t]
    \centering
    \includegraphics[width=0.49\textwidth]{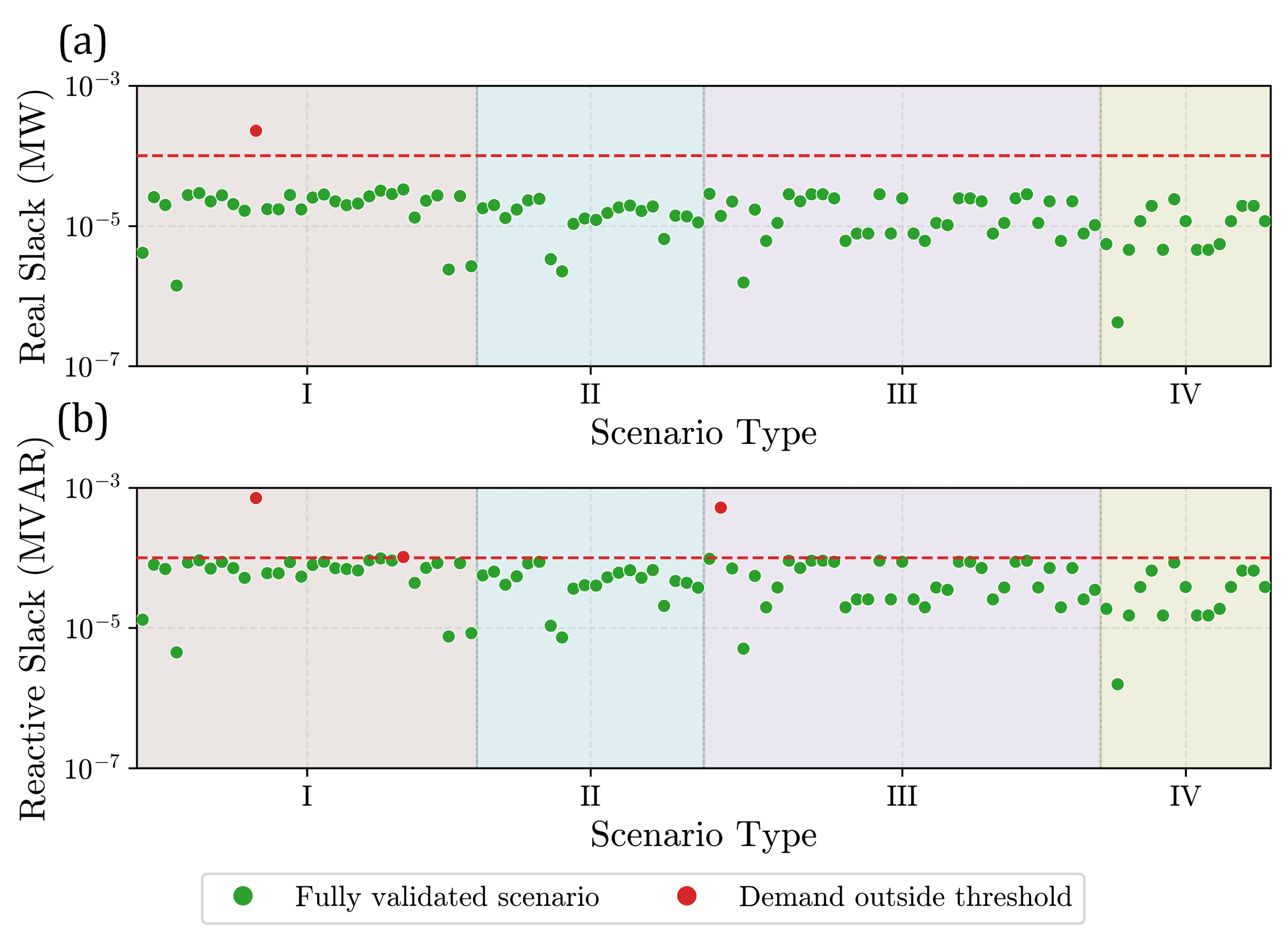}
    \caption{The results of the validation runs on the proposed microgrid design are shown, with (a) representing the total real slack $\sum_{t \in \mathcal{T}} \sum_{j \in \mathcal{N}} s^{\text{p}}{j,t}$ and (b) representing the total reactive slack $\sum{t \in \mathcal{T}} \sum_{j \in \mathcal{N}} s^{\text{q}}_{j,t}$ for each scenario. The different scenario categories are indicated using varying shades, and the red dotted line in each plot denotes the threshold value $\alpha = 10^{-4}$, above which slack values are considered invalid.}
    \label{fig:Validation}
    \vspace{-0.5cm}
\end{figure}

With the proposed microgrid capacities defined, we solve the validation problem in \eqref{eq:operational_problem_val} using loads from different scenarios. A threshold value of $\alpha = 10^{-4}$ is imposed, and all scenarios with total slack below this threshold are considered valid. To determine this threshold, we use the primal feasibility tolerance of the optimization solver, which in our case on the order of $10^{-6}$ for the MOSEK solver \cite{mosek_parameters}. Then, accounting for the accumulation of numerical residuals across constraints, we scale this tolerance is scaled by the number of constraints per scenario to obtain the slack threshold. In our case, each run consists of 288 time steps, and the constraint set is evaluated at each time step. This aggregation leads to an effective tolerance in the range of the selected $\alpha=10^{-4}$. The total slack for real and reactive power injections are shown in Fig.~\ref{fig:Validation}(a) and Fig.~\ref{fig:Validation}(b) respectively. The results indicate that the proposed microgrid performs satisfactorily in an average of 98\% of the scenarios, demonstrating robustness under varying traffic conditions, including extreme cases.

\subsection{Cost Comparison With and Without Traffic-Aware Microgrid Planning}\label{subsubsec: System Planning - With and Without Traffic Flow}
Throughout this work, we have emphasized the importance of coupling traffic flow models with grid planning, showing how DWC operational characteristics affect system design. With the microgrid now modeled, we quantify this impact by comparing costs with a design by assuming worst-case traffic for the roadway segment. While costs could also be compared to a grid designed for average demand (slightly lower), such a system cannot meet a large fraction of observed demand scenarios, as it ignores spatiotemporal demand patterns, as illustrated in Section~\ref{sec: Motivating Example}. Therefore, we focus the comparison on the worst-case design.

For this comparison, we use the median load we found in Section~\ref{subsubsec: Traffic Scenario Simulations}. The corresponding load peaks at about 28.33 MW, leading to a maximum energy demand of 2361~kWh on the charging coils over a 5-minute time step. Thus, we assume a flat load of 28.3 MW throughout the day, producing a worst-case microgrid capable of meeting this maximum at all times. In the traffic-aware case, we use the actual median spatiotemporal demand profile, capturing a more accurate representation of the system’s energy requirements.
\begin{figure}[t]
    \centering
    \includegraphics[width=0.48\textwidth]{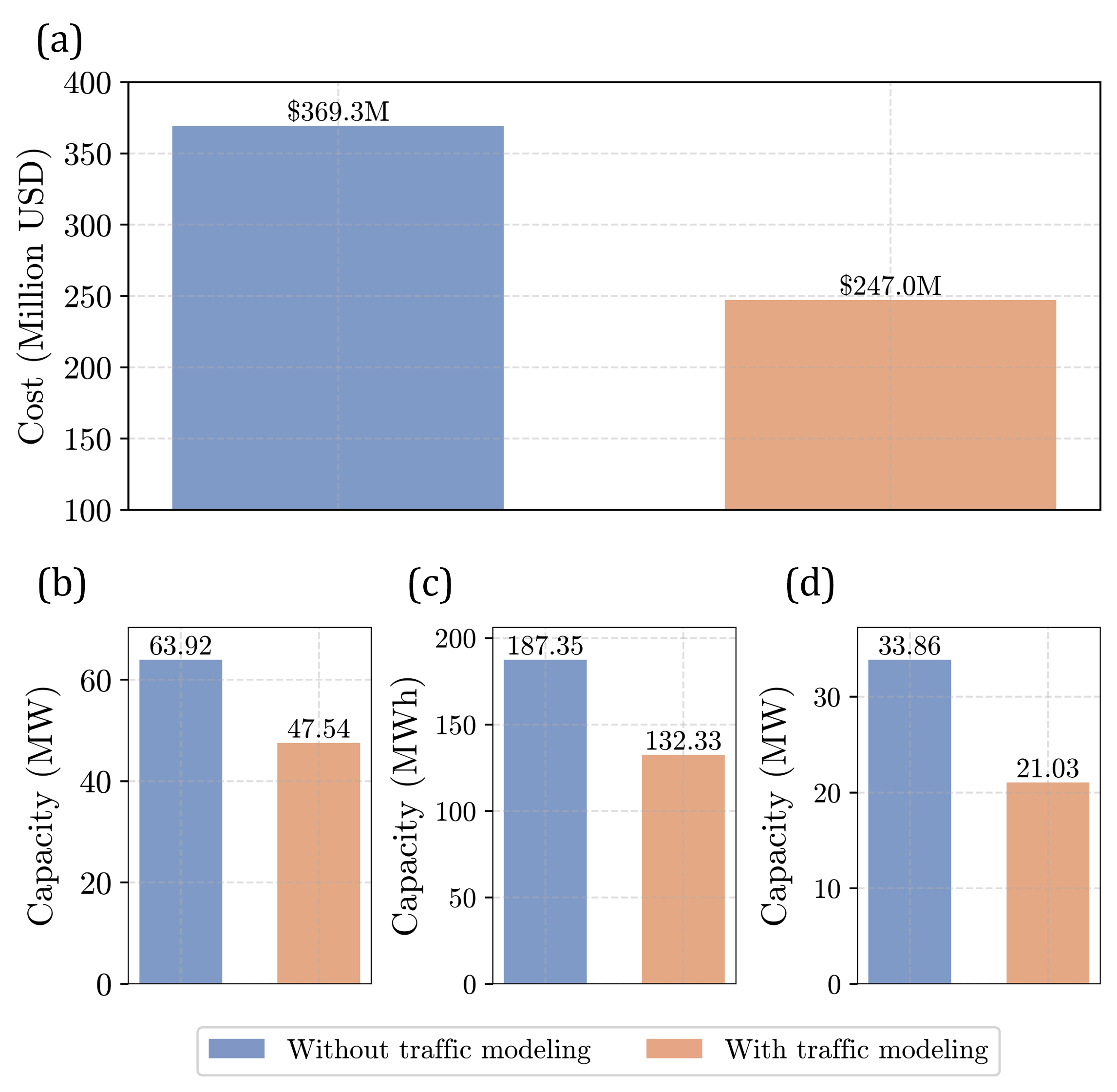}
    \caption{Comparing the system differences considering both traffic-aware and continuous worst case demand modeling approaches with: (a) showing the differences in overall system cost, (b) showing the differences in the installed solar capacity, (c) giving differences in the total ES capacity, and (d) showing the total grid coupling needed for both the systems. We see significant reduction in system size as well as overall cost when considering traffic-aware microgrid planing.}
    \label{fig:traffic_no_traffic_cost}
    \vspace{-0.5cm}
\end{figure}
As shown in the cost comparison in Fig.~\ref{fig:traffic_no_traffic_cost}, ignoring traffic-aware modeling and using a worst-case load profile leads to significant over-allocation of system costs. The total cost rises from approximately \$247 million in the traffic-aware case to about \$369.3 million when traffic dynamics are ignored, representing a 50\% increase. This difference in system sizing extends to every component of the proposed microgrid, as also shown in Fig.~\ref{fig:traffic_no_traffic_cost}. The optimal solar capacity rises from 47.5~MW to 63.9~MW (35\% increase), ES capacity grows from 132.33~MWh to 187.35~MWh (41.5\% increase), and external grid coupling changes from 21~MW to 33.86~MW (61\% increase) to meet the flat peak load. This demonstrates that neglecting traffic variations in microgrid planning leads to oversizing across all components, increasing capital and operational costs.
While peak traffic demand is important for system sizing, it occurs only briefly during the day and can be managed using a combination of generation resources, such as ES units that store energy during low-demand periods. Traffic-aware planning leverages this flexibility, aligning system design with realistic traffic-driven demand, resulting in a more optimally sized system for DWC operations and further reinforcing the utility of our planning framework.

\section{Discussion}\label{sec: Discussion}

This work addresses the challenges of integrating DWC with power grid planning. We propose a model that couples macroscopic traffic flows with  roadway energy demand from DWC coils and use this model to design a supporting microgrid. The system is validated across multiple traffic scenarios, with the final design ensuring reliable power under all realistic conditions. This traffic-aware microgrid planning avoids overdesign, reduces costs, and eliminates the need for vehicle-level monitoring by relying on models of average roadway behavior. The framework provides projected system size and investment requirements, offering stakeholders a practical tool for evaluating and implementing DWC along roadways.
 Note that our modeled microgrid parameters depend on assumed capital and operational costs. While we use the latest discretized costs adjusted for inflation, these can be updated based on the roadway region and construction year. The generalized mathematical formulation of our models and optimization framework allows adaptation to market trends and region-specific constraints, making them useful for a broad range of planners and investors.

Finally, we utilize the spatiotemporal variability of traffic demand to enhance system flexibility, enabling potential applications in ancillary services such as demand response. Future work will focus on leveraging the spatiotemporal variability of traffic demand to enhance system flexibility, enabling demand repsponse and participation in ancillary markets, further solidifying the role of DWC in sustainable grid operation.

\section{Acknowledgments}\label{sec: Acknowledgments}
This research was partially supported by a grant from the US National Science Foundation (NSF) Engineering Research Center on Advancing Self-sufficiency through Powered Infrastructure for Roadway Electrification (ASPIRE).

\balance
\bibliography{cas-refs}
\end{document}